\documentclass[conference]{IEEEtran}
\IEEEoverridecommandlockouts
\usepackage{cite}
\usepackage{amsmath,amssymb,amsfonts}
\usepackage{algorithm}
\usepackage{algorithmic}
\usepackage{graphicx}
\usepackage{textcomp}
\usepackage{xcolor}
\usepackage{makecell}
\usepackage{hyperref}
\hypersetup{
    colorlinks=true,
    linkcolor=blue,     
    citecolor=blue,     
    filecolor=blue,     
    urlcolor=blue       
}
\def\BibTeX{{\rm B\kern-.05em{\sc i\kern-.025em b}\kern-.08em
    T\kern-.1667em\lower.7ex\hbox{E}\kern-.125emX}}
\begin{document}

\title{KANOP: A Data-Efficient Option Pricing Model using Kolmogorov–Arnold Networks\\
}

\author{\IEEEauthorblockN{ Rushikesh Handal}
\IEEEauthorblockA{
\textit{Preferred Networks, Inc.}\\
Tokyo, Japan \\
rhandal@preferred.jp}
\and
\IEEEauthorblockN{Kazuki Matoya}
\IEEEauthorblockA{
\textit{Preferred Networks, Inc.}\\
Tokyo, Japan \\
kmatoya@preferred.jp}
\and
\IEEEauthorblockN{Yunzhuo Wang}
\IEEEauthorblockA{\textit{Preferred Networks, Inc.} \\
Tokyo, Japan \\
0000-0001-5843-482X}
\and
\IEEEauthorblockN{Masanori Hirano}
\IEEEauthorblockA{
\textit{Preferred Networks, Inc.}\\
Tokyo, Japan \\
research@mhirano.jp}
}

\maketitle

\begin{abstract} 

    Inspired by the recently proposed Kolmogorov–Arnold Networks (KANs), we introduce the KAN-based Option Pricing (KANOP) model to value American-style options, building on the conventional Least Square Monte Carlo (LSMC) algorithm. KANs, which are based on Kolmogorov-Arnold representation theorem, offer a data-efficient alternative to traditional Multi-Layer Perceptrons, requiring fewer hidden layers to achieve a higher level of performance. By leveraging the flexibility of KANs, KANOP provides a learnable alternative to the conventional set of basis functions used in the LSMC model, allowing the model to adapt to the pricing task and effectively estimate the expected continuation value. Using examples of standard American and Asian-American options, we demonstrate that KANOP produces more reliable option value estimates, both for single-dimensional cases and in more complex scenarios involving multiple input variables. The delta estimated by the KANOP model is also more accurate than that obtained using conventional basis functions, which is crucial for effective option hedging. Graphical illustrations further validate KANOP’s ability to accurately model the expected continuation value for American-style options.
    
\end{abstract}

\begin{IEEEkeywords}
Option Pricing, Least Square Monte Carlo, Kolmogorov–Arnold Networks, Basis Functions, option delta
\end{IEEEkeywords}

\section{Introduction}
Options are fundamental financial instruments that offer investors flexibility in hedging, speculation, and enhancing portfolio returns. Accurate option pricing is critical for market participants and financial institutions, as it aids in managing risk and identifying arbitrage opportunities. Equally significant is the precise calculation of delta, a key metric that measures an option's sensitivity to changes in the underlying asset's price. Both accurate option pricing and reliable delta calculations are essential for effective risk management, allowing market participants to construct and adjust portfolios that respond optimally to market fluctuations.

For European-style options, the Black-Scholes model proposed in \cite{b1} provides a closed-form solution, allowing for the calculation of both the option price and the Greeks associated with vanilla options. However, path-dependent options do not always enjoy this simplicity, and the absence of a closed-form solution makes their pricing more challenging. The Least Square Monte Carlo (LSMC) approach proposed in \cite{b2} offers a method to calculate option prices for these path-dependent options. The core idea behind LSMC is to approximate the expected continuation value using a set of basis functions and fitting this relationship via ordinary least squares (OLS). 

In theory, for the LSMC price to converge to the true option price, both the number of simulated paths and the degree of the polynomial basis functions must approach infinity \cite{b3}, which is impractical in real-world scenarios. Although the LSMC method is versatile and applicable to a variety of path-dependent options, it lacks clear guidelines for selecting the type and order of basis functions that ensure convergence to the true option price with a limited number of simulation paths. Inaccurate fitting of the expected future payoff can lead to incorrect option pricing and delta calculations, which are crucial for effective hedging.

The recently proposed Kolmogorov-Arnold Network (KAN) in \cite{b4} offers a promising alternative to alleviate the ambiguity in selecting the appropriate set of basis functions. KAN is inspired by the Kolmogorov-Arnold Representation Theorem (KART), which states that any multivariate continuous function on a bounded domain can be represented through a finite composition of continuous single-variable functions combined via binary addition. Unlike Multi-Layer Perceptrons (MLPs), which employ linear mappings between layers, KANs leverage learnable one-dimensional splines. This combination of splines and MLPs enables KANs to achieve spline-like accuracy, outperforming the linear functions of MLPs while avoiding the curse of dimensionality associated with traditional spline methods.

Due to their greater flexibility within each layer, KANs can learn the mapping from inputs to outputs without the need for a deep architecture, i.e., without increasing the number of hidden layers. This reduces the requirement for a large number of simulated paths to achieve high accuracy. Since the mapping from inputs, such as the underlying stock price, to the expected continuation value is typically smooth, KANs are particularly well-suited for learning this relationship. Moreover, the adaptability of hidden layer weights to the data further reduces reliance on specific basis function choices.

In this work, we propose KANOP, a KAN-based LSMC option pricing method, and address the central research question: \textbf{To what extent does KANOP provide accurate option value and delta estimates compared to conventional basis function-dependent LSMC algorithms under a limited number of simulated paths?}

To the best of our knowledge, we are the first to apply KANs to the option pricing task in the financial domain. The main contributions of our work are as follows:
\begin{itemize}
    \item We demonstrate that KANOP offers significantly more accurate price estimates for American and Asian options compared to conventional basis function-based LSMC models and deep MLP models.
    \item We further show that this improved accuracy extends to delta calculations, which is essential for effective hedging.
    \item Through graphical illustrations, we show that KANOP accurately fits the future expectation of the option payoff, leading to more precise price and delta calculations.
\end{itemize}

\section{Related Work}

Since inception, LSMC-based methods have gained popularity in the option pricing domain for their simplicity. Although being a versatile methodology, the choice of basis function in LSMC has been a topic of research for a long time. The foundational approach in \cite{b2} uses Laguerre polynomials as the basis function and also mentions Hermite, Legendre, Chebyshev, or Jacobi polynomials as potential candidates. For convergence, they suggest increasing the order of the polynomial until the value calculated by the LSMC algorithm does not increase. However, with an increase in polynomial order, the number of simulated paths also needs to be increased, making the approach computationally expensive. 

Various choices for basis functions, such as simple polynomials, splines, piecewise linear sparse grids, and sparse polynomials, are explained in \cite{b5}, which finds increased numerical efficiency for sparse polynomial functions. A similar approach of using polynomial sparse grid basis functions is implemented in \cite{b6} to price moving window Asian options. Recently, Hermitian basis functions have been utilized in a hierarchical tensor format for pricing Bermudan options in high-dimensional settings, as demonstrated in \cite{b7}. An LSMC-based method has also been applied for option pricing within the energy sector in \cite{b8}, where they conclude that the choice of basis function affects pricing when the option is not deep in-the-money (ITM) or out-of-the-money (OTM). Contrary to these methods, our learnable approach alleviates the need to explicitly specify the type of the basis function used.

In recent years, there has been a growing body of research focused on pricing methods that leverage machine learning (ML) techniques. The LSMC method itself is similar to Q-learning, a type of Reinforcement Learning (RL) algorithm, where the action is optimal stopping and the Q-values are updated via backward regression. Instead of full-scale RL-based approaches, several researchers have attempted to replace the linear mapping of LSMC with an ML-based model. For instance, a neural network (NN) model is used to find optimal stopping timing for Bermudan options in \cite{b9}. A similar NN-based method is employed for pricing and hedging American options in \cite{b10}. Additionally, \cite{b11} implemented support vector machine-based regression as a potential alternative for linear regression in LSMC. Contrary to these methods, our research offers the first study into the potential of KANs to replace the linear mapping part in LSMC-based methods.

Another popular approach for option pricing is based on solving partial differential equations (PDEs). The foundational work in \cite{b1} introduces the Black-Scholes PDE, which provides a framework for pricing European options. In the recent years, there has been an increase in research solving parabolic PDEs using the connection to Backward Stochastic Differential Equations (BSDEs), with \cite{b21} and \cite{b20}, being among the first few. Both classical numerical methods (see \cite{b12}, \cite{b13}, \cite{b14}, \cite{b15} for example) and ML-based methods (see \cite{b16}, \cite{b17}, \cite{b18} for example) are used to solve these BSDEs for option pricing. Unlike BSDE-based option pricing methods, which require elaborate formulations, LSMC-based methods are simpler and more versatile for various option structures.

Our research leverages the versatility of the LSMC methodology alongside the flexibility of KANs to propose a model capable of pricing a variety of options and addressing hedging problems.

\section{Methodology}
In the case of American-style execution options,  determining the conditional expectation of future payoffs is imperative; since it is optimal to exercise the option early if the immediate payoff, i.e., the intrinsic value of the option, exceeds the future expected payoff. With simulated price paths under a given market model, the LSMC method utilizes the cross-sectional information in the simulation, transformed with basis function mapping, to determine this value. Reliance only on a market model for price path simulation and the OLS method for parameter estimation renders LSMC versatile and straightforward to apply for option pricing.

For the sake of consistency, we adhere to the notations defined in \cite{b2}. In this framework, $(\Omega, \mathcal{F}, P)$ represents the complete probability space, where $\Omega$ denotes the set of all possible states from time 0 to $T$, with $T$ being a finite time horizon. An element $\omega \in \Omega$ represents a typical price path under some market model, while $\mathcal{F}$ and $P$ denote the sigma field and probability measure, respectively. The augmented filtration generated by the market model is defined as $F = \{\mathcal{F}_{t}; t \in [0, T] \}$ with $\mathcal{F}_{T} = \mathcal{F}$, and the risk-neutral measure is denoted $\mathbb{Q}$.

The focus of this research is options with American-style execution, i.e., they can be optimally executed at some point in $[0, T]$. To ensure no-arbitrage, the option price is equal to the minimum supermartingale that is greater than or equal to the discounted cash flows of the option, considering all potential stopping times within the filtration $\mathcal{F}$. The cash flow path generated by following this optional stopping strategy, conditioned on not being executed early at time $t$, is denoted by $C(\omega, s; t, T)$ where $t < s \leq T$. LSMC aims to approximate this optimal stopping time strategy by comparing the conditional expectation of $C(\omega, s; t, T)$ with the intrinsic value of the option, indicated by $V_{i}(\omega, t)$. 

At option maturity ($t = T$), the cash flow is equal to the intrinsic value of the option, $V_{i}(\omega, T)$. At each discrete time step before expiry $t_k$ with $0 < t_1 \leq t_2 \leq \cdots \leq t_k \leq \cdots \leq t_K = T$ (where $K$ denotes the number of discrete time steps), the expected value of continuation $F(\omega; \ t_k)$ is given by:
\begin{equation}
\mathbb{E}_{\mathbb{Q}} \left[ \sum_{j=k+1}^{K}  e^{-\int_{t_k}^{t_j}r(\omega, s)ds} C(\omega, t_j; t_k, T) \mid \mathcal{F}_{t_k} \right],
\label{eq:1}
\end{equation}

where $r(\omega, s)$ is the interest rate. With this value for continuation, LSMC follows a simple execution strategy to execute if $V_{i}(\omega, t_k)$ is positive and larger than $F(\omega; \ t_k)$. Since the cashflow at time $T$ is known, LSMC works backward in time, i.e., from $t_{K-1}$ to $t_1$ to approximate $F(\omega; \ t_k)$ using a linear combination of $\mathcal{F}_{t_k}$ measurable basis functions. For American-style options with payoff belonging to the $L^2$ square-integrable space, this approximation is justifiable, as $L^2$ being a Hilbert space, has a finite orthonormal basis. The coefficients for this linear combination are calculated using OLS estimation.

\subsection{Existing KAN Model}
Based on the KART, the recently proposed KANs offer a promising alternative to more traditional MLPs. The theorem states that a multivariate continuous function on a bounded domain can be expressed as a finite composition of univariate functions combined by addition. Following \cite{b4}, the mathematical representation of the theorem is given as:

\begin{equation}
    f(\textbf{x})\ = f(x_1, x_2, \dots, x_n) = \sum_{i=1}^{2n+1} \phi_i \left( \sum_{j=1}^{n} \psi_{ij}(x_j) \right),
    \label{eq:2}
\end{equation}

where \( f(\textbf{x}) \) \(( [0,1]^n \to \mathbb{R} )\) is a continuous function of \( n \) dimensional variable \textbf{x} with \( x_1, \ldots, x_n \) as elements, \( \psi_{ij} \)   \(( [0,1] \to \mathbb{R} )\) and \( \phi_i \) \(( \mathbb{R} \to \mathbb{R} )\) are continuous univariate functions. The representation expresses \( f \) as a sum of composed functions of the form \( \phi_i(\cdot) \).

With \( F(\omega; \ t_k) \) given in \eqref{eq:1} generally being continuous function of the underlying variables, KART can be used for approximate estimation.  For example, in the case of an American call option with a given strike price, \( F(\omega; \ t_k) \) is a smoothly increasing function of the stock price. 

A major advantage over conventional basis function formulations is that using cross-products of the input variables is unnecessary. Notice that \( \phi_i \) and \( \psi_{ij} \) given in \eqref{eq:2} are univariate functions of the input variables, and interactions among different variables occur only through the binary operation of addition. This simplicity presents significant potential for valuing options in high dimensions, where LSMC methods require higher-order basis functions and complex interactions between input variables, thereby increasing model complexity and necessitating a larger number of simulated paths to ensure reasonable convergence.

A drawback of KART is that it does not specify the type of function suitable for \( \phi_i \) or \( \psi_{ij} \), which in practice could be non-smooth or even fractal, making KART challenging to use for functional approximation. A recent breakthrough was introduced with the KAN implementation in \cite{b4}, where additional hidden layers are incorporated, in contrast to the two-layer model in KART. The KAN implementation parameterizes the univariate functions in this deeper and wider model using B-spline curves, with the coefficients of the B-spline basis functions being learnable. 

By increasing the number of hidden layers, KANs with smooth univariate functions can approximate even non-smooth \( \phi_i \) functions from KART. Thus, KANs offer the flexibility of splines to fit univariate functions, combined with the learnability of MLPs, enabling similar accuracy to MLPs but with a smaller network. This is particularly useful when option valuation is performed under a limited number of simulated paths.

\subsection{Proposed KAN-based Option Pricing Model}
In this work, we introduce the KANOP model, which integrates the flexibility of KANs with the versatility of the LSMC methodology. At each discrete time $t_k$, we assume that the expectation of the future payoff, $F(\omega; t_k)$, can be approximated using a KAN model with an appropriate layer structure. Since this expectation function is generally smooth, KART accommodates this assumption. 

At each time step, we initialize a new KAN network to fit $F(\omega; t_k)$, allowing for finer control over the functional form. Using the trained KAN model, we compute the fitted values of this expectation, $\hat{F}(\omega; t_k)$, and the option is exercised early if $V_{i}(\omega, t_k)$ is positive and larger than $\hat{F}(\omega; t_k)$. Starting from time $t_{K-1}$, where $C(\omega, t_K; t_{K-1}, T)$ is known, we iteratively fit the future expectation value and evaluate early exercise, moving backwards in time from $t_{K-1}$ to $t_1$. The option price $V_0$ is then given by the average of discounted cashflows over all paths $\Omega$. Algorithm \ref{alg:kanop} outlines the entire process for pricing an American-style option.

\begin{algorithm}
\caption{KAN-based Option Pricing (KANOP)}
\label{alg:kanop}
\begin{algorithmic}[1]
    \STATE \textbf{Input:} Price paths $\Omega$, Discrete time steps $t_k$ with $0=t_0< t_1 \leq t_2 \cdots t_k \leq\cdots t_K = T$
    \STATE \textbf{Initialize:} Terminal cashflow $C(\omega, t_K; t_{K-1}, T)$
    \FOR{$k = K-1$ \textbf{to} 1}
        \STATE Calculate the discounted cashflow as  $  \sum_{j=k+1}^{K}  e^{-\int_{t_k}^{t_j}r(w, s)ds} C(\omega, t_j; t_k, T)$
        \STATE Approximate $F(\omega; \ t_k)$ using KAN
        \STATE Calculate fitted values $\hat{F}(\omega; \ t_k)$ 
        \STATE Early execute option if $V_{i}(\omega, t_k)$ is positive and greater than $\hat{F}(\omega; \ t_k)$ 
        \STATE Update $\{C(\omega, t_j; t_{k-1}, T)\}_{j=k}^{K}$ 
    \ENDFOR
    
    \STATE \textbf{Output:} Option price $ V_0$ as average of discounted cashflows over all paths $\Omega$ 
\end{algorithmic}
\end{algorithm}

For approximation using KAN, we employ the Mean Squared Error between $F(\omega; t_k)$ and $\hat{F}(\omega; t_k)$ as the loss function, which is consistent with the loss function used in OLS estimation. Compared to conventional LSMC approach, the overall algorithm remains the same, with the only difference being that the linear combination of basis functions is replaced by a learnable KAN model. This approach constitutes a general pricing algorithm suitable for a broad range of American-style options, with modifications only needed for the calculation of the intrinsic value function $V_{i}(\omega, t_k)$. For a standard American option, this is the difference between the underlying stock price and the strike price. For an Asian option with American-style exercise, it is the difference between the average stock price and the strike price. This simplicity in the LSMC model contributes to its versatility in option pricing.

\subsection{Proposed Method for Approximate Delta Calculation}
Delta $\Delta$ measures the sensitivity of an option's price to small changes in the price of the underlying asset. Mathematically, delta is defined as the partial derivative of the option's price with respect to the underlying asset price $S_0$, given by:
\begin{equation}
    \Delta = \frac{\partial V_0}{\partial S_0}.
    \label{eq:3}
\end{equation}

Delta hedging is a strategy employed to manage the risk associated with fluctuations in the price of the underlying asset. The objective of delta hedging is to maintain a delta-neutral portfolio, ensuring that the portfolio’s overall exposure to small changes in the underlying asset price is zero.

Since the LSMC algorithm calculates the point estimate of the option price rather than the entire functional form around $S_0$, using \eqref{eq:3} to calculate delta is not feasible. However, LSMC provides an alternative method for delta calculation that leverages simulated paths. The value $\hat{F}(\omega; t_1)$ represents the expected continuation value at time step $t_1$, assuming the option is not exercised at $t_1$. We define the value for a path at $t_1$, $V(\omega, t_1)$, as:

\begin{equation}
    V(\omega, t_1) = \max\{\hat{F}(\omega; \ t_1), V_{i}(\omega, t_1)\}
    \label{eq:4}
\end{equation}

If the model can accurately approximate $F(\omega; t_1)$ for all paths, then $\hat{F}(\omega; t_1)$ provides a reliable cashflow estimate for all paths discounted to $t_1$. Consequently, the option value is given by the discounted mean of $V(\omega, t_1)$, expressed as:

\begin{equation}
    V_0 = \mathbb{E}_{\mathbb{Q}} \left[e^{-\int_{0}^{t_1}r(w, s)ds}V(\omega, t_1)\mid \mathcal{F}_{0} \right].
    \label{eq:5}
\end{equation}

Delta can be computed by taking the derivative of this value with respect to the stock price at $t_0$. We utilize Autograd, a powerful tool that enables automatic computation of gradients with respect to model inputs. This approach ensures that if the model accurately approximates $F(\omega; t_1)$, it will also yield an accurate delta estimate. By focusing on paths up to time $t_1$ rather than all paths up to time $T$, the delta calculation becomes numerically efficient. Since a reliable approximation of $F(\omega; t_1)$ is required for both ITM and OTM paths, we use the entirety of the generated simulated paths to fit ${F}(\omega; t_1)$. To address the added complexity in fitting, we also double the number of basis functions used in conventional LSMC, as suggested in \cite{b2}.

To illustrate the applicability of the proposed method for delta calculation, consider a European call option with a $S_0$ of $100$, a strike price $K_p$ of $102$, a time to maturity (TTM) of $30$ days (discretized at daily frequency), an annual volatility $\sigma_y$ of $20\%$, an annual risk-free rate $r_y$ of $0\%$, and an annual dividend rate $\delta_y$ of $0\%$. 

In this case, the true form of $F(\omega; t_1)$ is simply the Black-Scholes price calculated with the corresponding price path having a TTM of $29$ days. We assume that $\hat{F}(\omega; t_1)$ can fit this functional form perfectly. Since there is no early exercise for a European option, $V(\omega, t_1)$ is equal to $\hat{F}(\omega; t_1)$. The delta calculated using the proposed approximation method with Autograd is $0.4008$, which matches the delta value of $0.4008$ obtained from the closed-form Black-Scholes formula for European options. This example validates the proposed approximation method for delta calculation using simulated paths within the LSMC framework.

\section{Experiments}

We illustrate the applicability of KANs for option pricing and hedging using two examples of American-style options: a standard American option, where the sole variable is the underlying stock price, and an Asian American option, where both the stock price and the time-weighted average price (TWAP) serve as variables. The American option offers insight into KAN pricing for the simpler, single-variable case, while the Asian American pricing task evaluates the model's performance as the number of underlying variables increases. A versatile model should deliver accurate price estimates in both scenarios.

\subsection{Standard American Option}\label{sec:exp_am}

For the standard American option experiment, offering the simplest example of an American-style option, we consider a put option with the following specifications:
\begin{center}
$S_0 = 4.0$, \\
$K_p = 4.0$, \\
$TTM\ T = 50\ days$, \\
$\sigma_y = 20\%$, \\
$r_y = \delta_y = 0\%$. \\
\end{center}
In the case of an American call option with no dividend, it is not optimal to exercise the option early since the time value of the American option is always positive. As for the American put option, even under no risk-free rate and zero dividend, it might be optimal to exercise the option early. Consider a case when the stock price drops to $0$, where the intrinsic value is the highest attainable, equal to the strike price. In such a case, it is optimal to exercise the put option early.

With the chosen values for $\sigma_y$ and TTM, simulated paths with a stock price of $0$ are non-existent. Consequently, in this simpler scenario, early exercise of the American put option is not optimal, and the option behaves similarly to a European put option with the same specifications. Therefore, the true form of ${F}(\omega; \ t_k)$ for each path $\omega$ is simply the European put option value $P(\cdot)$, calculated for the given path with price $x$ at time $t_k$, and TTM taken as $T - t_k$; given as:

\begin{equation}
\label{eq:bs}
\begin{aligned}
P(x, K_p, t_k, r_{t_k}, \sigma_{t_k}) &= K_p e^{-r_{t_k} (T-t_k)} \Phi(-d_2) - x \Phi(-d_1), \\
d_1 &= \frac{\ln\left(\frac{x}{K_p}\right) + \left(r_{t_k} + \frac{\sigma_{t_k}^2}{2}\right)(T-t_k)}{\sigma_{t_k} \sqrt{(T-t_k)}}, \\
d_2 &= d_1 - \sigma_{t_k} \sqrt{(T-t_k)},
\end{aligned}
\end{equation}
where $\sigma_{t_k}$ is the square-root adjusted volatility and $r_{t_k}$ is the linearly adjusted risk-free rate for the TTM $T - t_k$. $\Phi(\cdot)$ is the cumulative distribution function of the standard normal distribution.

For all LSMC-based models, we use only $10,000$ simulated paths, consistent across all models. This relatively small simulated path size tests each model's accuracy in price estimation under limited data. The time to maturity is discretized at a daily frequency, allowing the option to be exercised early each day. Technically, this makes it a Bermudan put option, but with no interest rate or dividend, the option behaves like a European put. Consequently, the price and delta of this option are equal to those obtained from the closed-form solution of the Black-Scholes model.

\subsubsection{Conventional LSMC Dependence on Basis Functions}
First, we assess whether the claim made in \cite{b2}, that the choice of basis function does not impact the calculated option price, holds true under a limited number of simulated paths. For this analysis, we utilize weighted Laguerre polynomials (Weighted Laguerre model) as presented in \cite{b2} and compare their performance against Hermite polynomials (Hermite model) as alternative basis functions. In both scenarios, we fit $F(\omega; t_k)$ over the entire set of simulated paths, including OTM options, and thus use the first six orders of the basis functions to accommodate the increased complexity of approximation. 

A general form of Laguerre polynomials of $p^{th}$ order, \( L_p(x) \), is given as:
\begin{equation}
\begin{aligned}
    L_0(x) &= 1, \\
    L_1(x) &= 1 - x, \\
    L_p(x) &= \frac{1}{p}[(2p - 1 - x) L_{p-1}(x) - (p - 1) L_{p-2}(x)].
\end{aligned}
\end{equation}

Weighted Laguerre polynomials simply weigh $ L_p(X) $ by $e^{-x/2}$. Using exponential weighting can result in computational underflow in cases of large values for $x$, but with moderate volatility and price in our example, this is not an issue.

Similarly, the general form of the Hermite polynomial of $p^{th}$ order, $H_p(x)$, is given as:
\begin{equation}
\begin{aligned}
    H_0(x) &= 1, \\
    H_1(x) &= 2x, \\
    H_p(x) &= 2x \cdot H_{p-1}(x) - 2(p-1) H_{p-2}(x).
\end{aligned}
\end{equation}

\subsubsection{LSMC Models utilizing Learnable Basis Functions}
In this experiment, we evaluate whether LSMC models that learn the basis functional mapping, such as the KANOP model or an MLP model, are more effective at approximating ${F}(\omega; \ t_k)$. Here, we use a simple KANOP model with the $[1,3,1]$ structure, featuring only one single hidden layer. Here, the $[1,3,1]$ structure denotes a network with a 1-dimensional input, a hidden layer with 3 units, and a 1-dimensional output. The number of units in the hidden layer adheres to the $(2n+1)$ rule from the original KART. The KANOP model utilizes $10,000$ simulated paths, which are identical to those employed for the Weighted Laguerre and Hermite models, ensuring a fair comparison in performance evaluation.

As an alternative approach for the learnable basis function model, we employ an MLP model that utilizes traditional neural network layers for inter-layer mapping. A challenge with using an MLP model without a deep layer structure is that such a shallow configuration offers limited flexibility to fit a complex functional form. Conversely, incorporating a deeper layer structure necessitates a substantial amount of data for effective training. This issue is also emphasized in \cite{b4}.

To ensure the MLP model has sufficient expressive power, we use a larger number of simulated paths; specifically, ten times more for training. This increase applies only to the MLP model, while KANOP uses just $10,000$ simulated paths, giving the MLP model a notable advantage in fitting ${F}(\omega; \ t_k)$. The MLP model for pricing the American option is structured as $[1, 32, 32, 1]$, trained on $100,000$ simulated paths, including those used in KANOP training. We apply Algorithm \ref{alg:kanop} for fitting the MLP model at each discrete time step until maturity.

\subsection{Asian American Option}\label{sec:exp_as}
In this experiment, we evaluation a model's performance when the number of underlying input variables increases. An American option requires estimating ${F}(\omega; \ t_k)$ using the stock price as the sole underlying variable. In the previous standard American option pricing example with a zero dividend and risk-free rate, any function that overestimates ${F}(\omega; \ t_k)$ will yield a correct option price. This is because, in scenarios where $\hat{F}(\omega; \ t_k)$ exceeds $V_{i}(\omega, t_k)$, early exercise will not occur. The Asian American option provides a case for comparing various models, especially when early exercise is optimal for some paths. Here, both the stock price and the TWAP are variables for basis function modeling.

For experimentation, We consider the Asian American option example described in \cite{b19}. Here, we assume a call option with $V_{i}(\omega, t_k)$ defined as the difference between the TWAP at time $t_k$ ($TWAP_{t_k}$) and the strike price $K_p$. With a TTM of $T_w$ weeks, the option can be exercised at the end of any week until $T_w$. Thus, in practice, the option is a Bermudan style option, with a discrete exercise time step $t_k$ representing the end of each week.

The example consists of four different option pricing tasks. The first task uses $S_0$ of $100$, $K_p$ of $100$, $T_w$ of $13$ weeks, $\sigma_y$ of $15\%$, and $r_y$ of $5\%$. The second task increases $\sigma_y$ to $25\%$, while the third task subsequently extends the TTM to $26$ weeks. The fourth task further raises $K_p$ from $100$ to $105$. This example includes at-the-money (ATM) options as well as OTM option to evaluate the model's performance in estimating ${F}(\omega; \ t_k)$.

Following \cite{b19}, we focus solely on calculating the option value for the Asian American option, unlike the American option pricing example, where we also compute delta. Furthermore, the availability of early execution enhances the option value for the Asian American option compared to the Asian option with the same specifications, which is executed in a European style, known as the Eurasian option. We utilize the values provided in \cite{b19} as the actual option prices for the Asian American option.

For the conventional LSMC model, following \cite{b2}, we employ Laguerre polynomials (Laguerre model) as the chosen basis functions. With no exponential weighting, the space spanned by the linear combination of Laguerre polynomials does not differ from that of the Hermite polynomials. To accommodate OTM paths, we apply transformations up to the fourth order of the polynomial evaluated at $S_t$ and $TWAP_t$, along with the cross products of these transformations. This results in a total of 15 regressors for OLS estimation, and we rely on limited $10,000$ simulated paths to estimate the corresponding parameters. 

The KANOP model utilizes the same simulated paths as the Laguerre model but increases the model size to a $[2, 5, 1]$ structure to accommodate two input variables. Here too, we adhere to the $2n+1$ rule for the number of units in the hidden layer. In contrast, the MLP model employs a size of $[2, 32, 32, 1]$, and for the MLP, we again utilize 10 times more simulated paths for fitting. The paths used for fitting the Laguerre and KANOP models are also included in the simulated path set for the MLP model.

\section{Results}
Here, we present findings from each experiment. First, we demonstrate the impact of basis function choices in the conventional LSMC algorithm on the pricing and hedging tasks for American options.  Next, we compare the performance of KANOP with that of the MLP and conventional models using the same tasks. Finally, we compare the Asian American option prices calculated by the KANOP model against those from the conventional LSMC model and the MLP.

\subsection{Standard American Option}

With the input values for the standard American option provided in Section \ref{sec:exp_am}, the option price calculated using the Black-Scholes model is $0.1421$.  The mean of $V_{i}(\omega, 50)$ across all $10,000$ paths is $0.1422$, which is close to the theoretical option price. Hence, we use the theoretical option price as the target option value. As the option is an ATM option, the corresponding put delta is $-0.5$.

The estimates for option values and deltas are summarized in Table \ref{tab:am}. Overall, the KANOP model demonstrates a superior combination of both price estimates and delta. We further provide illustrations for the sub-experiments for the standard American option in the following subsections.

\begin{table}[htbp]
\caption{Estimates for American put option price and delta with $S_0=4.0$, $K_p=4.0$, $TTM=50$ days, $\sigma_y= 20\%$ and $r_y=\delta_y=0\%$. Black-Scholes price and delta are the target values.}
\label{tab:am}
\begin{center}
\begin{tabular}{|c|p{1.5cm}|p{1.5cm}|}
\hline
\textbf{Model} & \makecell{\textbf{\textit{Price}}} & \makecell{\textbf{\textit{Delta}}} \\
\hline
Black-Scholes Model & \makecell{0.1421} &  \makecell{-0.5000} \\
\hline
Weighted Laguerre Model & \makecell{0.1395}  & \makecell{ -0.4876 }\\
Hermite Model & \makecell{0.1407} &  \makecell{-0.4899} \\
MLP Model & \makecell{0.1384} &  \makecell{\textbf{-0.4976}} \\
KANOP Model (Ours)& \makecell{\textbf{0.1427}} &  \makecell{-0.4970} \\
\hline
\end{tabular}
\end{center}
\end{table}

\subsubsection{Conventional LSMC Dependence on Basis Functions}

\begin{figure}[!t]
    \centering
    
    \begin{minipage}[b]{\columnwidth}
        \centering
        \includegraphics[width=0.8\columnwidth]{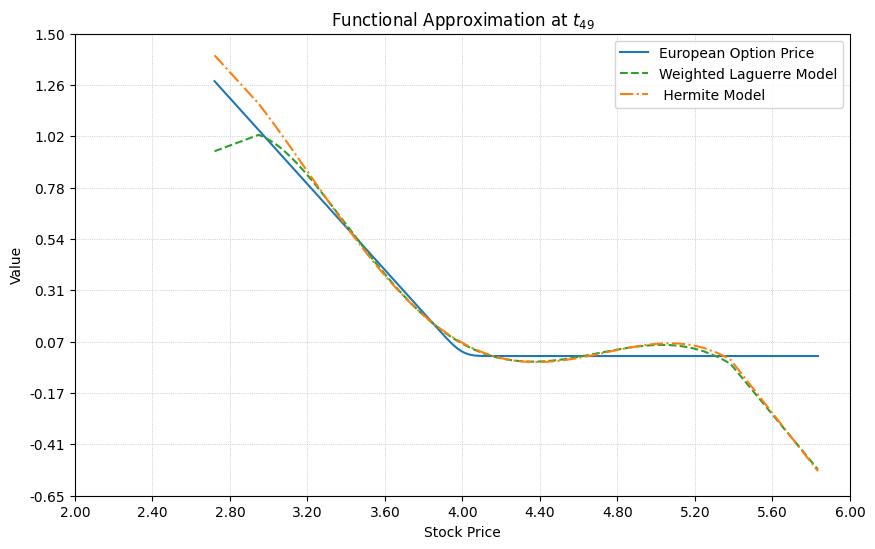}
        \label{fig:plot1}
    \end{minipage}
    
    \vspace{1em}
    
    \begin{minipage}[b]{\columnwidth}
        \centering
        \includegraphics[width=0.8\columnwidth]{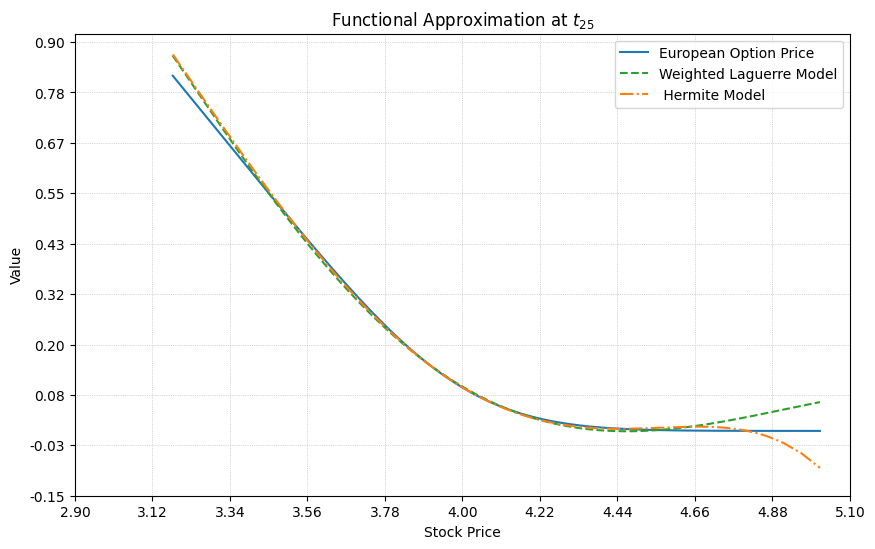}
        \label{fig:plot2}
    \end{minipage}
    
    \vspace{1em}
    
    \begin{minipage}[b]{\columnwidth}
        \centering
        \includegraphics[width=0.8\columnwidth]{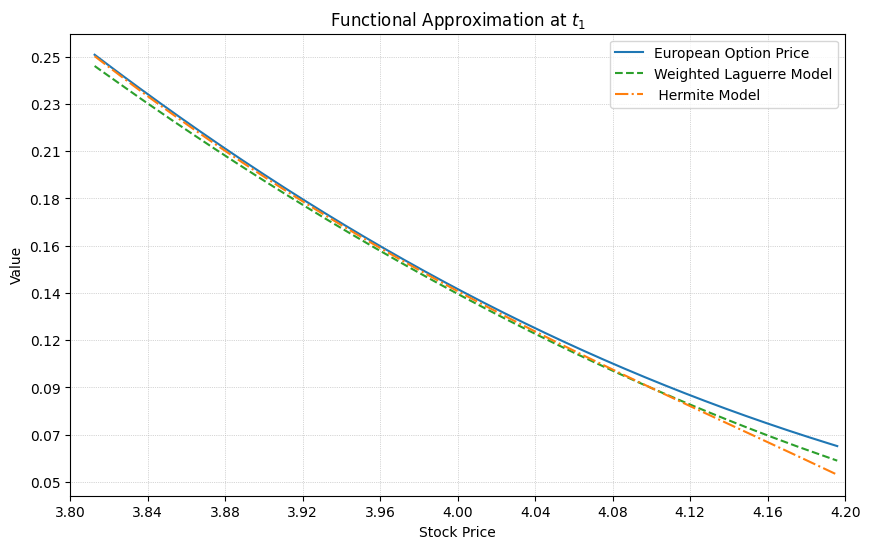}
        \label{fig:plot3}
    \end{minipage}
    
    \caption{$\hat{F}(\omega; \ t_k)$ values using the Weighted Laguerre model and Hermite model; indicating at $t_1$, Hermite model model provides a better approximation.}
    \label{fig:lh}
\end{figure}

The values of $\hat{F}(\omega; \ t_k)$ approximated at discrete times $t_{49}$, $t_{25}$, and $t_1$ for the Weighted Laguerre model and Hermite model are shown in Fig.~\ref{fig:lh}. Overall, the approximated $\hat{F}(\omega; \ t_k)$ is inaccurate for both sets of basis functions when the path is either deep ITM or deep OTM. At $t_{49}$, the values for $\hat{F}(\omega; \ t_{49})$ near the strike price are overestimated by these models. However, these inaccuracies do not directly propagate to previous time steps, and both models can fit ${F}(\omega; \ t_{25})$ reasonably well, with the Hermite model performing slightly better than the Weighted Laguerre model.

A similar argument applies to estimating ${F}(\omega; \ t_1)$, where the Hermite model performs better than the Weighted Laguerre model for ITM paths. Consequently, the Weighted Laguerre model yields a price estimate of $0.1395$, resulting in an error of $1.81\%$, while the price estimated by the Hermite model is $0.1407$, with an error of $0.97\%$, nearly half that of the counterpart. This simple example demonstrates that, with a limited number of generated simulated paths, the choice of basis function significantly impacts the option price estimate.

The delta estimated by the Weighted Laguerre model is $-0.4876$, while the actual value is $-0.5$. In contrast, the delta estimated by the Hermite model is $-0.4899$, again showing slight improvement over the Weighted Laguerre model.

\subsubsection{LSMC Models utilizing Learnable Basis Functions}

\begin{figure}[!t]
    \centering
    
    \begin{minipage}[b]{\columnwidth}
        \centering
        \includegraphics[width=0.8\columnwidth]{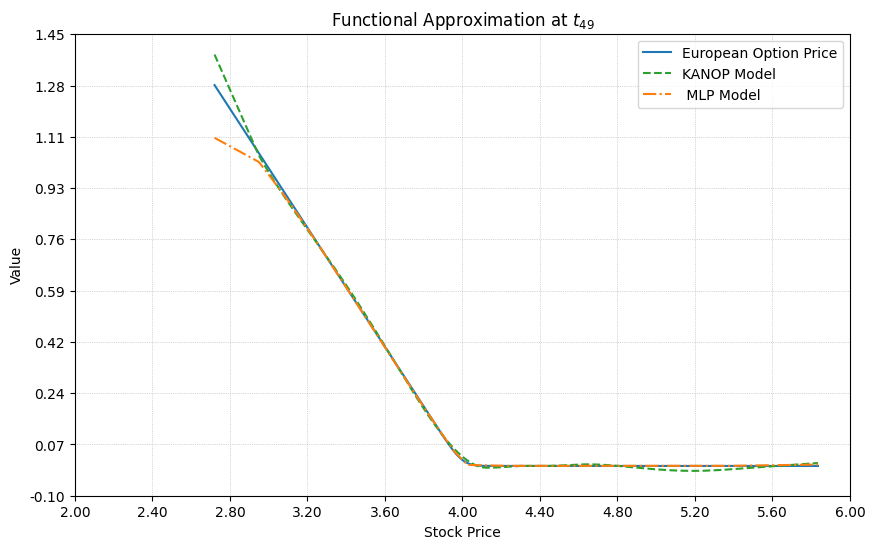}
        \label{fig2:plot21}
    \end{minipage}
    
    \vspace{1em}
    
    \begin{minipage}[b]{\columnwidth}
        \centering
        \includegraphics[width=0.8\columnwidth]{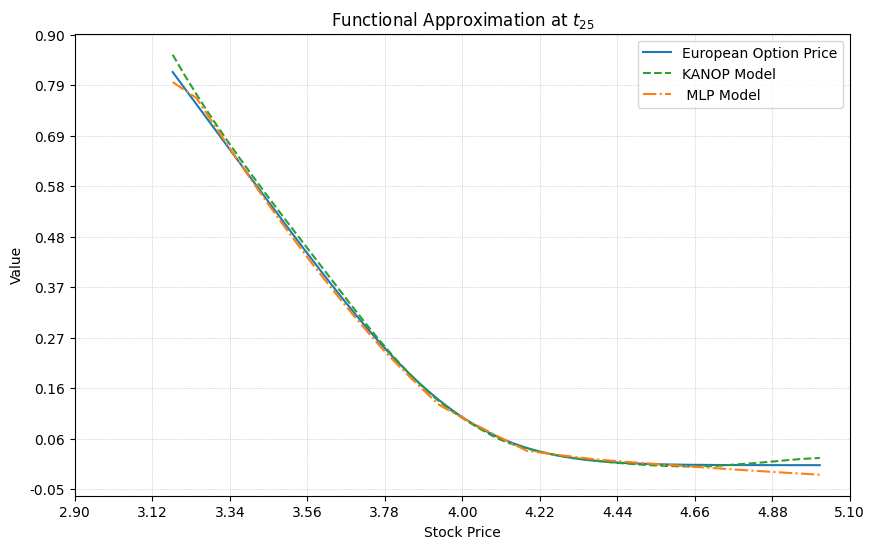}
        \label{fig2:plot22}
    \end{minipage}
    
    \vspace{1em}
    
    \begin{minipage}[b]{\columnwidth}
        \centering
        \includegraphics[width=0.8\columnwidth]{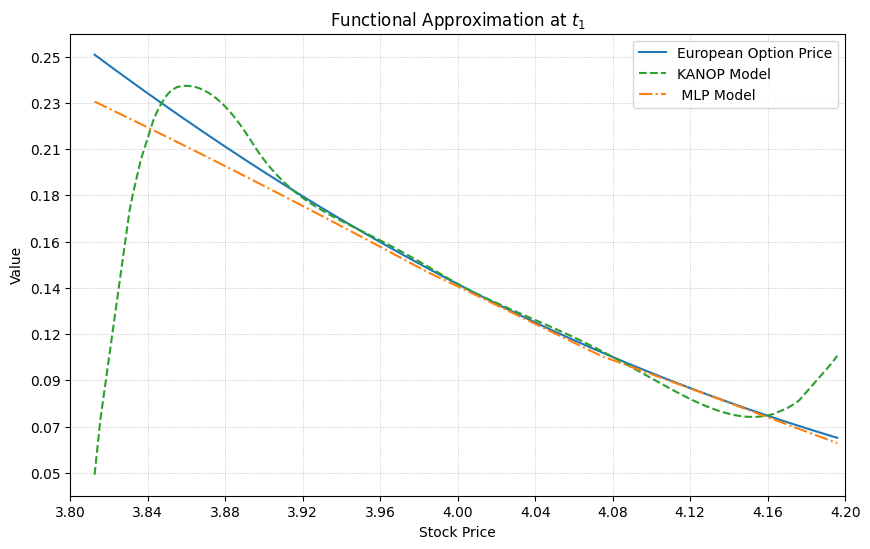}
        \label{fig2:plot23}
    \end{minipage}
    
    \caption{$\hat{F}(\omega; \ t_k)$ values using the KANOP model and MLP model; indicating at $t_1$, KANOP model provides a better approximation.}
    \label{fig:km}
\end{figure}

Similarly, the corresponding approximations of $\hat{F}(\omega; \ t_k)$ at discrete times $t_{49}$, $t_{25}$, and $t_1$ for the KANOP model and the MLP model are shown in Fig.~\ref{fig:km}. A notable distinction from the Weighted Laguerre and Hermite models is that both the KANOP and MLP models can approximate ${F}(\omega; \ t_{49})$ significantly better. Particularly at stock prices near $K_p$, where the time value for the remaining day is the dominant component of the option value, both learnable basis function models provide accurate estimates.

At time step $t_{25}$, both the models approximate the European option price quite well. However, in the case of the MLP, this accuracy does not carry over to the estimate of ${F}(\omega; \ t_1)$, resulting in an underestimation of the expected continuation value. The corresponding option price calculated by the MLP is $0.1384$, which is even less accurate than that of the Weighted Laguerre model.

In the case of the KANOP model, the approximation near $K_p$ is significantly better than that of the MLP model, closely aligning with the actual ${F}(\omega; \ t_1)$. The larger deviations observed around extreme prices at time $t_1$ are attributed to the smaller number of paths; and hence, these deviations do not significantly impact the price or delta estimates. The price estimated by the KANOP model is $0.1427$, resulting in a difference of only $0.44\%$ compared to the actual price. The deltas estimated by the MLP and KANOP models are $-0.4976$ and $-0.4970$, respectively, both showing an improvement over conventional basis function models.

\subsection{Asian American Option}

For the Asian American option experiments with the settings given in Section \ref{sec:exp_as}, the estimated option values for different models are summarized in Table \ref{tab:as}. Asian American and Eurasian option prices are provided for comparison. For the Eurasian prices, we calculate the average of the discounted payoffs from time $T_w$ using the same generated $10,000$ simulated paths; common across all models. In all cases, the target Asian American prices exceed those of the Eurasian prices, indicating the benefit of early exercise.

\begin{table}[htbp]
\caption{Estimates for Asian American call option price for given $K_p$, $T_w$ \& $\sigma_y$. Eurasian prices are given for comparison. KANOP model provides the best price estimates for all combinations.}
\label{tab:as}
\begin{center}
\begin{tabular}{|c|c|c|c|c|}
\hline
&  {\textit{$K_p$=100}} &  {\textit{$K_p$=100}} &  {\textit{$K_p$=100}} &  {\textit{$K_p$=105}} \\
&  {\textit{$T_w$=13}} &  {\textit{$T_w$=13}}&  {\textit{$T_w$=26}}&  {\textit{$T_w$=26}} \\
 {Model} &  {\textit{$\sigma_y$=0.15}}  &  {\textit{$\sigma_y$=0.25}}  &  {\textit{$\sigma_y$=0.25}}  &  {\textit{$\sigma_y$=0.25}} \\
\hline
Eurasian Price &  2.1638 &   3.3621 &   4.7659&    2.6628 \\

Asian American Price &  2.3210 &  3.6500&   5.2660&  2.8580 \\
\hline
Laguerre Model & 2.2750 & 3.5716 & 5.0719 & 2.7162\\
MLP Model & 2.2601 & 3.6134  &5.1422 & 2.7943 \\
KANOP Model (Ours)& \textbf{2.3216} & \textbf{3.6589} & \textbf{5.2382} & \textbf{2.8309} \\
\hline
\end{tabular}
\end{center}
\end{table}

The Laguerre model provides price estimates that are higher than the Eurasian prices but still lower than the actual Asian American prices. This discrepancy is particularly pronounced for the OTM option. This suggests that the Laguerre model leads to inaccurate estimations for ${F}(\omega; \ t_k)$, resulting in early executions occurring at inappropriate times or no early execution when it would be optimal.

The MLP model, in contrast, provides comparable or improved value estimates relative to the Laguerre model. Despite utilizing 10 times more simulated paths than the other models, the MLP price estimates still fall short of the actual Asian American option values. However, the KANOP model, even with a limited number of simulated paths, yields superior option value estimates compared to both the models. The discrepancies between the KANOP model's estimates and the actual Asian American prices are within just a few cents. 

The Laguerre model lacks the flexibility to adjust the complexity of its basis functions based on data, while the MLP model struggles with small simulated path sizes. In contrast, the KANOP model, featuring only one hidden layer with 5 units, is able to approximate ${F}(\omega; \ t_k)$ significantly better.

\section{Discussion}

For any LSMC-based option pricing model, the estimate $\hat{F}(\omega; \ t_k)$ is only used to determine early execution. A given path triggers an early execution if $V_{i}(\omega, t_k)$ positive; and exceeds $\hat{F}(\omega; \ t_k)$. Thus, for option pricing, an imprecise estimate $\hat{F}(\omega; \ t_k)$ is acceptable as long as its directional relationship to $V_{i}(\omega, t_k)$ (higher or lower) aligns with that of the true expected value of continuation ${F}(\omega; \ t_k)$. Significant deviations in $\hat{F}(\omega; \ t_k)$ compared to ${F}(\omega; \ t_k)$ are inconsequential for OTM paths,  $V_{i}(\omega, t_k)$ is negative.

In the case of the standard American option example, the true form of ${F}(\omega; \ t_k)$, represented by the European put value in Equation \eqref{eq:bs}, involves a logarithmic function combined with $\Phi(\cdot)$. For the conventional LSMC model, a linear combination of basis function polynomials must approximate the non-linearity of this expression. Evident from Fig.~\ref{fig:lh}, both the Weighted Laguerre model and the Hermite model struggle to accurately fit ${F}(\omega; \ t_{49})$. This is  particularly true where value changes are sharper near the strike price, indicating the inefficiency of simple linear combinations. In contrast, both learnable basis function models effectively approximate ${F}(\omega; \ t_{49})$; as can be seen in Fig.~\ref{fig:km}.

Since $\hat{F}(\omega; \ t_k)$ is used solely to determine early execution, the inaccuracies of the conventional models in fitting ${F}(\omega; \ t_{49})$ do not necessarily impact performance at earlier time steps. Both of the conventional models are able to fit ${F}(\omega; \ t_{25})$ reasonably well. However, as the fitting progresses backward in time, the Weighted Laguerre model underestimates ${F}(\omega; \ t_{1})$, leading to incorrect early execution timings and resulting in a poorer price estimate compared to the Hermite model; evident from Fig.~\ref{fig:lh} as well as Table \ref{tab:am}.

As can be seen in Fig.~\ref{fig:km} and Table \ref{tab:am}, a similar argument applies to the MLP model, which leads to an underestimation of the option price even with $10$ times more simulated paths. The KANOP model, on the other hand, struggles to fit ${F}(\omega; \ t_{1})$ in the deep ITM or deep OTM regions due to the limited number of paths in these areas. However, it provides a correct estimate for most of the ITM paths, resulting in a more accurate option price.

Contrary to the option value, the actual value of $\hat{F}(\omega; \ t_1)$ apart from the direction is crucial for delta calculation as it determines the expected value of future cash flows for a given path.  Visually, as seen in Fig.~\ref{fig:lh} and Fig.~\ref{fig:km}, the curvature of $\hat{F}(\omega; \ t_1)$ at the stock price equal to $K_p$ provides a clear indication of delta, since the slope at $t_1$ represents the delta with one less day until maturity. As a result, KANOP offers a better delta estimate from both of these perspectives.

For the Asian American option example, increasing the number of input variables while keeping the number of simulated paths constant adds difficulty for a model to fit ${F}(\omega; \ t_k)$. Evident from Table \ref{tab:as}, the conventional Laguerre model fails to fully capture the option value. This indicates that Laguerre model leads to sub-optimal early execution for some simulated paths, stemming from incorrect estimation of ${F}(\omega; \ t_k)$. A similar argument applies to the MLP model as well. In contrast, KANOP, with its local flexibility using splines and a learnable approach for tuning, provides a more accurate price estimate compared to its counterparts.

\section{Conclusion}
In this research, we utilize the recently developed KAN model and introduce the KANOP model to answer the central question of this study: To what extent does KANOP provide accurate option value and delta estimates compared to conventional basis function-dependent LSMC algorithms under a limited number of simulated paths?

Using an American put option example, we demonstrate that, under a limited number of simulated paths, the choice of basis functions affects the estimated option price and delta in the case of conventional LSMC. The KANOP model provides a learnable alternative to these conventional basis functions, resulting in a better estimation of the option value as well as delta. Using graphical illustrations, we show that KANOP excels at correctly approximating the expected value of continuation.

To achieve a reasonable approximation for the expected value of continuation, MLP models typically require a deeper structure, necessitating a larger number of simulated paths. In contrast, we show that the KANOP model with just one hidden layer provides more accurate option prices, both in the standard American option example as well as the Asian American pricing task. In the Asian American pricing example as well, KANOP leads to much smaller errors in price estimates compared to conventional basis functions used in the LSMC method.

With the consistent performance of KANOP for pricing single-dimensional American options and the Asian American option with two dimensions, KANOP emerges as a promising choice for pricing higher-dimensional options, such as rainbow options or basket options, providing a foundation for future research.


\end{document}